\documentclass[aps,twocolumn,pra,showpacs,floatfix]{revtex4}
\usepackage{epsfig}
\usepackage{graphicx}
\usepackage{dcolumn}

\begin{document}


\title{Correlation potential and ladder diagrams}

\author{V. A. Dzuba}
\affiliation{School of Physics, University of New South Wales,
Sydney 2052, Australia}

\date{\today}

\begin{abstract}

The all-order correlation potential method of accurate atomic structure 
calculations for atoms with one external electron is extended to include one
more class of correlation diagrams to all orders. These are the so-called
{\em ladder} diagrams which describe residual Coulomb interaction between an external
electron and atomic core. This is in addition to the screening
of Coulomb interaction by core electrons and the hole-particle interaction 
in the core polarization operator which are also included in all orders.
Calculations of the energies of the lowest $s$, $p$ and $d$ states of
cesium and thallium show that inclusion of the ladder diagrams leads
to significant improvement of the accuracy of the calculations. 
The discrepancy between theoretical and experimental energies is reduced
to a small fraction of a per cent in all cases. This widens the range of 
atoms and atomic states for which the correlation potential method can
produce very accurate results.

\end{abstract}


\maketitle

\section{Introduction}

There are many areas in modern physics which require accurate atomic 
calculations.
This includes parity and time invariance violation in atoms~\cite{Guena,ginges},
search for variation of the fundamental constants~\cite{Flambaum07a,Lea,Uzan}, 
atomic clocks~\cite{Hall,Hansch}, etc. Calculations are needed for planing
of the experiments and for interpretation of the results.
Accuracy of atomic calculations is often a limitation factor.
For example, the most accurate measurements of the parity
non-conservation (PNC) in atoms was done in Boulder in 1997
for the cesium atom~\cite{Wood}.
The accuracy of the measurements is 0.35\%. 
Theoretical accuracy of best calculations is on the level of 
0.4 - 0.5\% and as a result the accuracy of extraction of the 
weak charge of the cesium nucleus is only 0.6\%~\cite{ginges}.
The situation is even worse for the PNC in thallium where best
experimental accuracy is 1\%~\cite{Fortson} while the accuracy of best
calculations is 3\%~\cite{Dzuba87} and 2.5\%~\cite{Kozlov}.
Since atomic PNC measurements serve as an important source of
information about low energy physics and possible extensions
to the standard model, further improvements in the accuracy of
atomic calculations is highly desirable.

In present paper we limit our discussion to monovalent atoms 
keeping in mind PNC in Cs and Tl and other similar important
applications. Ground state configuration of thallium is 
[Xe]$5d^{10}6s^26p$ and to some extend it can be treated 
as an atom with one external electron above the [Xe]$5d^{10}6s^2$
closed-shell core. This approach was used in our early
calculations of the PNC in Tl~\cite{Dzuba87}. It is generally
believed however that for more accurate results Tl should
be treated as an atom with three external electrons above the
[Xe]$5d^{10}$ closed-shell core. This is because the $6s$
electrons are easy to excite which is evident from the
existence of the states in thallium discrete spectrum which
belong to the $6s6p^2$ configuration. A method which
combines the configuration interaction (CI) technique 
for valence electrons with the many-body perturbation theory
(MBPT) for the core-valence correlations was suggested in 
Ref.~\cite{CI+MBPT}. Recent calculations of the Tl PNC performed
with the use of this method \cite{Kozlov} achieved 
only moderate improvement of accuracy,
from 3\% in Ref.~\cite{Dzuba87} to 2.5\% in Ref.~\cite{Kozlov}.

In present paper we advocate a different approach which treats 
thallium atom as a monovalent system but includes dominating 
classes of core-valence correlations in all orders. It is 
based on the all-order correlation potential method which
was developed in Ref.~\cite{Dzuba89} and used in a number
of calculations mostly for alkali-metal atoms~\cite{Dzuba89a,Dzuba89b,
Dzuba95,Dzuba01,Dzuba02}. To make it work equally well for other
atoms like thallium we extend the technique to include one
more class of higher-order diagrams, the {\em ladder} diagrams.
This diagrams describe residual Coulomb interaction of the external
electron with the core. The idea of the extension is inspired by 
the coupled-cluster (CC) approach. This is another powerful method
which is widely used for monovalent atoms. In this approach, the 
many-electron wave function of an atoms is written in terms of
single, double, etc. excitations from the reference Hartree-Fock
wave function. The accuracy depends on the number of terms included
into expansion and limited by available computer power. The method 
was used by many groups for accurate calculations of wide range of
properties of many-electrons atoms (see, e.g. Ref.~\cite{CC})
including PNC in Cs~\cite{Blundell90}. There are plans to use this
approach to improve the accuracy of calculations of the PNC in Cs
by including more terms into the expansion~\cite{Derevianko07}.

Although the CC approach can produce very accurate results it is 
computationally very demanding. Even relatively simple singe-double
approximation takes significant computer resources but insufficiently
accurate for some atoms, e.g. heavy alkali-metal atoms~\cite{Johnson07}.
In contrast, the all-order correlation potential method is very
efficient. However, it produces accurate results only for $s$ and
$p$ states of alkali-metal atoms. The accuracy for other monovalent atoms 
and for the $d$ states of alkali-metal atoms is significantly lower. In general,
the overlap between wave functions of the valence and core electrons
must be small for accurate results. Small overlap would mean small
residual Coulomb interaction between valence and core electrons.
To overcome this limitation of the all-order correlation potential
method we use the CC-like equations to include 
residual Coulomb interaction between valence and core electrons in
all orders. Corresponding terms in the MBPT are presented by ladder
diagrams.

To test the technique we calculate lowest $s$, $p$ and $d$ energy
levels of cesium and thallium. Consideration of the lowest states
is sufficient for testing of the calculations of the correlations.
This is because correlations are smaller for excited states and
within the same technique accuracy of calculations is usually
better for excited states. Therefore, wide range of the different
states of monovalent atoms are covered.

We demonstrate that the inclusion of the ladder diagrams leads to 
significant improvements in the accuracy of calculation.
This opens a way of atomic structure calculations for many 
important applications with the accuracy which was not available before.

\section{Correlation potential}

\label{CPM}

\begin{figure}
\centering
\epsfig{figure=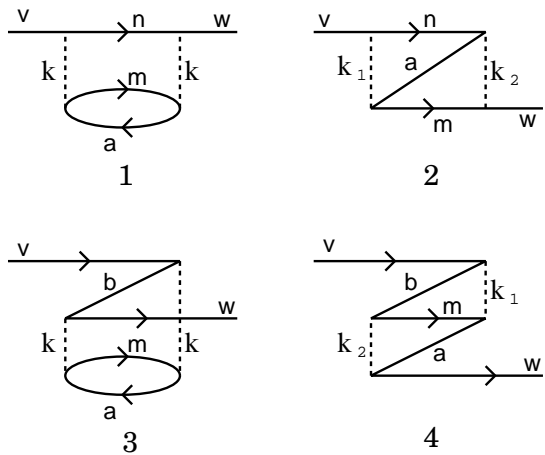,scale=0.8}
\caption{Second order correlation diagrams for $\hat \Sigma$}
\label{sigma1}
\end{figure}

The all-order correlation potential method was developed in Refs.~\cite{Dzuba89}
and successfully used for a number of calculations for alkali-metal atoms and their
isoelectronic ions. The method is based on the use of the so-called
correlation potential $\hat \Sigma$ which is defined in such a way that its
expectation value over a wave function $|v\rangle$ of a valence electron is 
equal to the many-body perturbation theory (MBPT) expression for the correlation 
correction to the energy of the electron
\begin{equation}
  \delta \epsilon_v = \langle v |\hat \Sigma | v \rangle.
\label{def-sigma}
\end{equation}
The correlation potential $\hat \Sigma \equiv \Sigma_v(r_1,r_2)$
is a non-local operator similar to the Hartree-Fock (HF) exchange potential.
It can be used in the HF equations for valence electrons to calculate the 
so-called Brueckner orbitals
\begin{equation}
(\hat H^{HF} +\hat \Sigma - \epsilon_v)\psi_v =0.
\label{Brueck}
\end{equation}
Here $\hat H^{HF}$ is the HF Hamiltonian. Solving the equation (\ref{Brueck})
for different states of external electron produces the wave functions and 
the energies which include correlations.

Following our earlier works~\cite{Dzuba89,Dzuba89a,Dzuba89b,Dzuba95,Dzuba01} 
we use the all-order correlation potential $\hat \Sigma^{(\infty)}$ which includes
two classes of higher-order correlations: (a) screening of Coulomb interaction between
a valence electron and a core electron by other core electrons and 
(b) an interaction between an electron excited from the core and a hole
created by this excitation. One more class of higher-order diagrams, 
iteration of the $\hat \Sigma$-operator, is included by iterating the 
equations for Brueckner orbitals (\ref{Brueck}).

The MBPT expansion for the correlation correction operator $\hat \Sigma$
starts from the second order. All four Brueckner-Goldstone diagrams are shown
on Fig.~\ref{sigma1} (to be more precise matrix elements
$\langle v|\hat \Sigma^{(2)}|w\rangle$ of the second-order
correlation potential $\hat \Sigma^{(2)}$ are shown). 
However, it is more convenient to use the Feynman diagram
technique to include dominating higher-order correlations. We do this for 
direct diagrams 1 and 3 on Fig.~\ref{sigma1}. Direct diagrams strongly
dominate over exchange ones in most of the cases and require accurate
treatment. The higher-order effects
for exchange diagrams (2 and 4 on Fig.~\ref{sigma1}) are included in
a semi-empirical way by introducing screening factors as it will be
explained below.

\begin{figure*}
\centering
\epsfig{figure=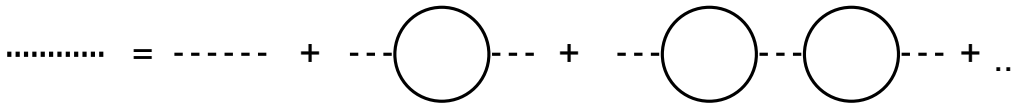,scale=0.7}
\caption{Screening of Coulomb interaction by polarization of the atomic core}
\label{screening}
\end{figure*}

\begin{figure*}
\centering
\epsfig{figure=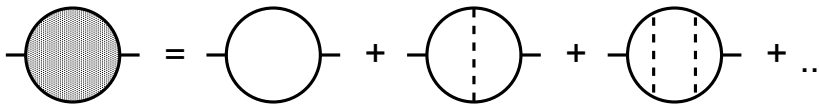,scale=0.7}
\caption{Hole-particle interaction in the polarization operator}
\label{hole-particle}
\end{figure*}

\begin{figure}
\centering
\epsfig{figure=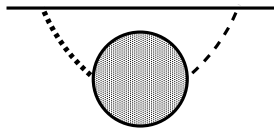,scale=0.7}
\caption{All-order correlation potential $\hat \Sigma$}
\label{sigma-all}
\end{figure}

Screening of Coulomb interaction are included by inserting core polarization
loops into Coulomb lines as shown on Fig.~\ref{screening}. Hole-particle 
interaction in the polarization operator is shown on Fig.~\ref{hole-particle}.
The all-order $\hat \Sigma$ operator is shown on Fig.~\ref{sigma-all}. This 
operator is drawn using Feynman diagram technique, screened Coulomb interaction
(Fig.~\ref{screening}) and the core-polarization operator with the hole-particle
interaction in it (Fig.~\ref{hole-particle}). This diagram does not include
exchange terms. Exchange diagrams are much smaller and screening for them
is included in a semi-empirical way via screening factors $f_k$. It is assumed
that screening depends only on the multipolarity of the Coulomb interaction $k$
and every Coulomb integral $g_k$ in diagrams 2 and 4 on Fig.~\ref{sigma1} is 
replaced by $f_kg_k$, where screening factors $f_k$ are found from the calculation 
of the direct diagram (Fig.~\ref{sigma-all}). The values of the screening
factors for Cs and Tl are given in Table~\ref{tb:fk}.

\begin{table}
\caption{Screening factors $f_k$ to calculate exchange diagrams of $\hat \Sigma$.}
\label{tb:fk}
\begin{ruledtabular}
  \begin{tabular}{cl lllllll}
Atom & State      & $f_0$ & $f_1$ & $f_2$ & $f_3$ & $f_4$ & $f_5$ & $f_6$    \\ \hline
Cs   & All        & 0.72  & 0.62  & 0.83  & 0.89  & 0.94  & 1.00  & 1.00 \\
Tl   & $7s_{1/2}$ & 0.54  & 0.55  & 0.90  & 0.89  & 0.95  & 0.97  & 0.99 \\
     & $6p_{1/2}$ & 0.71  & 0.67  & 0.85  & 0.90  & 0.95  & 0.97  & 0.99 \\
     & $6p_{3/2}$ & 0.74  & 0.58  & 0.86  & 0.89  & 0.97  & 0.97  & 0.99 \\
     & $6d_{3/2}$ & -.19  & 0.54  & 0.90  & 0.92  & 0.97  & 0.98  & 0.99 \\
     & $6d_{5/2}$ & 0.14  & 0.55  & 0.90  & 0.92  & 0.97  & 0.98  & 0.99 \\
  \end{tabular}
\end{ruledtabular}
\end{table}

\begin{table}
\caption{Energies of the lowest $s$, $p$ and $d$ states of Cs and Tl
in different approximations (cm$^{-1}$); comparison with experiment.}
  \label{tb:CP}
  \begin{ruledtabular}
    \begin{tabular}{cc rrrrr}
Atom & State    &  RHF  &
$\Sigma^{(2)}$\footnotemark[1] &
$\Sigma^{(\infty)}$\footnotemark[2] & \multicolumn{1}{c}{$\Delta$\footnotemark[3]} & 
 Exp.\cite{Moore} \\
\hline
Cs & $6s_{1/2}$ & 27954 & 32377 & 31462 &  -55 & 31407 \\
   & $6p_{1/2}$ & 18791 & 20523 & 20296 &  -67 & 20229 \\
   & $6p_{3/2}$ & 18389 & 19927 & 19728 &  -53 & 19675 \\
   & $5d_{3/2}$ & 14138 & 17459 & 17166 & -258 & 16908 \\
   & $5d_{5/2}$ & 14163 & 17305 & 17050 & -240 & 16810 \\
		       		 	       	 	
Tl & $7s_{1/2}$ & 21109 & 23375 & 22887 & -101 & 22786 \\
   & $6p_{1/2}$ & 43823 & 51597 & 50815 &-1551 & 49264 \\
   & $6p_{3/2}$ & 36636 & 43524 & 42491 &-1020 & 41471 \\
   & $6d_{3/2}$ & 12217 & 13428 & 13296 & -150 & 13146 \\
   & $6d_{5/2}$ & 12167 & 13319 & 13160 &  -96 & 13064 \\
    \end{tabular}
\footnotetext[1]{Brueckner orbitals with the second-order $\hat \Sigma$}
\footnotetext[2]{Brueckner orbitals with the all-order $\hat \Sigma$}
\footnotetext[3]{$\Delta = {\rm E}_{\rm exp} - {\rm E}_{\rm calc}(\hat \Sigma^{(\infty)}$}
  \end{ruledtabular}
\end{table}

The results of the calculations for Ca and Tl with the all-order correlation potential
method are presented in Table~\ref{tb:CP}. The final results are very accurate for the 
$s$ and $p$ states of Cs and $s$ and $d$ states of Tl. In general, the all-order
correlation potential method gives very accurate results for systems in which
external electron is on large distances from the atomic core and its residual
Coulomb interaction with the core is small. This is because this interaction is
included in the second-order of the MBPT only. This remains true even when all three
classes of the higher-order diagrams discussed above are included. Second
order is insufficient for cases of large overlap between wave functions of the 
core and the valence electron. This is the case for the ground state of Tl 
due to large overlap between the $6s$ and $6p$ states as well as for 
the $5d$ states of Cs due to large overlap between the $5d$ and $5p$ states. 
Here one needs to include residual Coulomb interaction 
between external electron and atomic core in all-orders to get accurate results. 
This can be done via the so called ladder diagrams which will be discussed in
next section.

\begin{figure*}
\centering
\epsfig{figure=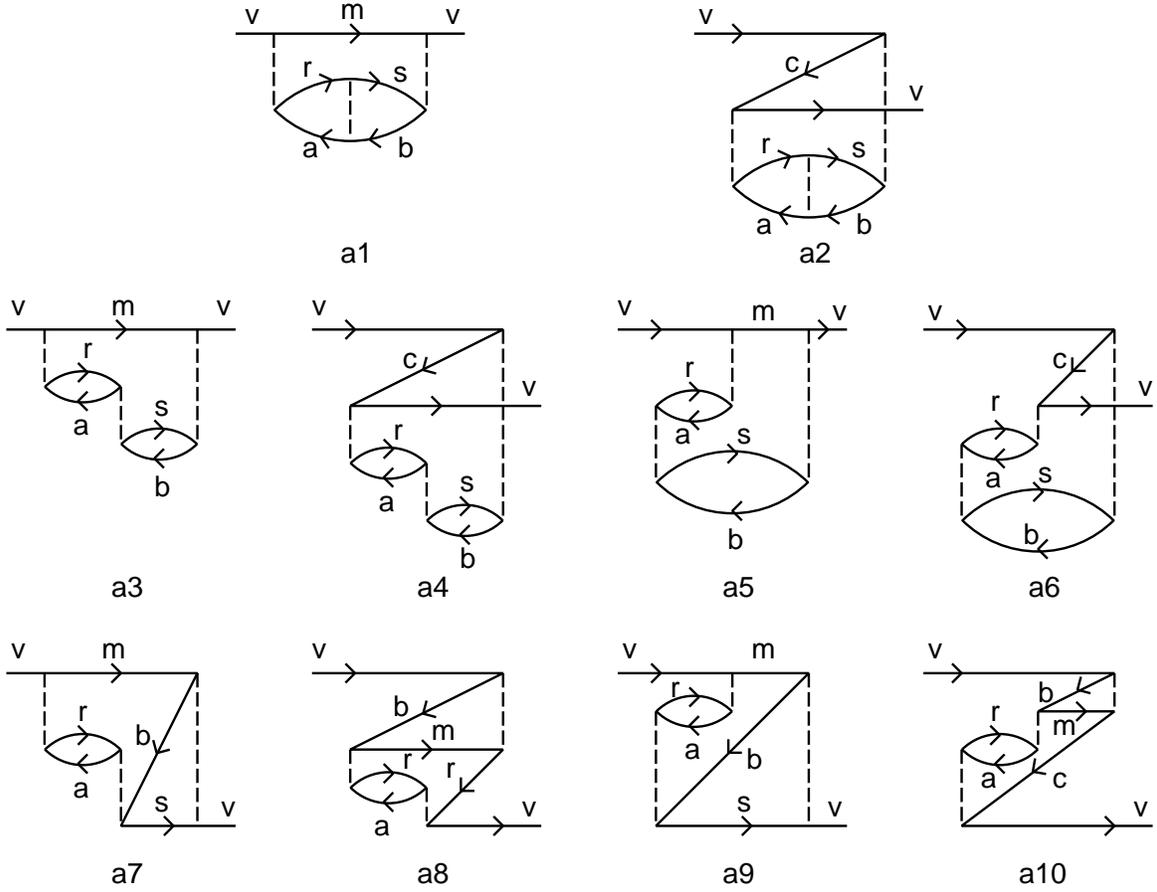,scale=0.8}
\caption{Third order diagrams corresponding to the correlation potential method.
Diagrams a1 and a2 include hole-particle interaction. Other diagrams include
screening of the Coulomb interaction. Diagrams a5-a10 have mirror-reflection partners.
Exchange diagrams are not shown.}
\label{cor3}
\end{figure*}

In the end of this section we present on Fig.~\ref{cor3} all third order diagrams
corresponding to the the all-order correlation potential method. The purpose is to
demonstrate that there is no overlap between the higher-order diagrams of the
all-order correlation potential method and the ladder diagrams of the next
section. No overlap means no double counting. 

\section{Ladder diagrams}

\begin{figure}
\centering
\epsfig{figure=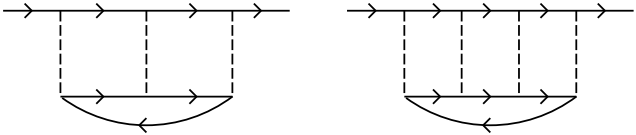,scale=0.6}
\caption{Sample third and forth order ladder diagrams.}
\label{l4}
\end{figure}

As it has been discussed in previous section, in atoms with large overlap
between the wave wave functions of the core and valence electrons the
residual Coulomb interaction between external electron and the core should
be included in all orders. This can be done via the so called ladder
diagrams. An example of the third and forth order ladder diagrams are shown on 
Fig.~\ref{l4}. In higher orders such diagrams have many parallel Coulomb lines
representing interaction of the external electrons with the core. This makes
a diagram to look like a ladder. Here is the name.

\begin{figure}
\centering
\epsfig{figure=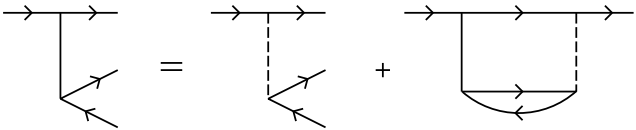,scale=0.6}
\caption{Graphic equation which generates ladder diagrams on Fig.~\ref{l4}.}
\label{leq}
\end{figure}

Direct calculation of the higher-order ladder diagrams is impractical. 
A much more efficient way is to perform an appropriate iteration procedure
in which each iteration corresponds to next order of the MBPT but takes
exactly the same time. For example, diagrams on Fig.~\ref{l4} can be
obtained by iteration the graphical equation of Fig.~\ref{leq}. Here
solid line represents an effective Coulomb interaction which is initially
equal to the ordinary Coulomb interaction and then adds a Coulomb line
to itself with every new iteration. Figs.~\ref{l4} and \ref{leq} do not
represent all ladder diagrams. They only show some higher-order extensions 
of the diagram 1 of Fig.~\ref{sigma1}. To include all ladder diagrams 
we must make sure that all Coulomb lines in all four diagrams of 
Fig.~\ref{sigma1} are repeated many times in the cause of iterations.
Extra Coulomb lines must be connected to the lines of electrons excited from
the core as well as to the lines of the holes created in the core by
electron excitations. For example, apart from diagrams on Fig.~\ref{l4}
there are must be companion diagrams in which arrows in the lower loop
go opposite direction.

The equations which satisfy these conditions can be written as two
sets of equations. The first is for atomic core:
\begin{eqnarray}
  && (\epsilon_a+\epsilon_b-\epsilon_m-\epsilon_n)\rho_{mnab} =
  g_{mnab}+ \label{lcore} \\ 
  && \sum_{rs}g_{mnrs}\rho_{rsab}+\sum_{rc}(g_{cnbr}\rho_{mrca}+
  g_{cmar}\rho_{nrcb}) \nonumber .
\end{eqnarray}
And another is for a specific state $v$ of an external electron:
\begin{eqnarray}
  && (\epsilon_v+\epsilon_b-\epsilon_m-\epsilon_n)\rho_{mnvb} =
  g_{mnvb}+ \label{lval} \\ 
  && \sum_{rs}g_{mnrs}\rho_{rsvb}+\sum_{rc}(g_{cnbr}\rho_{mrcv}+
  g_{cmvr}\rho_{nrcb}) \nonumber .
\end{eqnarray}
Here parameters $g$ are Coulomb integrals 
\[ g_{mnab} = \int \int \psi_m^\dagger(r_1) \psi_n^\dagger(r_2)e^2/r_{12}
\psi_a(r_1)\psi_b(r_2)d\mathbf{r}_1d\mathbf{r}_2, \] 
variables $\rho$ are
the coefficients representing expansion of the atomic wave function 
over double excitations from the zero-order Hartree-Fock reference 
wave function; parameters $\epsilon$ are the single-electron Hartree-Fock
energies. Coefficients $\rho$ are to be found by solving the equations
iteratively starting from
\[ \rho_{mnij} = \frac{g_{mnij}}{\epsilon_i+\epsilon_j-\epsilon_m-\epsilon_n}. \]
Indexes $a,b,c$ numerate states in atomic core, indexes
$m,n,r,s$ numerate states above the core, indexes $i,j$ numerate
any states.

The equations for the core (\ref{lcore}) do not depend on the 
valence state $v$ and are iterated first. The convergence is controlled
by the correction to the core energy
\begin{equation}
  \delta E_C = \frac{1}{2}\sum_{abmn} g_{abmn}\tilde\rho_{mnab},
\label{deltaec}
\end{equation}
where
\[ \tilde\rho_{mnab} = \rho_{mnab} - \rho_{mnba}. \]

When iterations for the core are finished the equations
(\ref{lval}) are iterated for as many valence states $v$ as needed.

Correction to the energy of the valence state $v$ arising from the
iterations of equations (\ref{lcore}) and (\ref{lval}) is given by
\begin{equation}
  \delta \epsilon_v = \sum_{mab}g_{abvm}\tilde\rho_{mvab}+
  \sum_{mnb}g_{vbmn}\tilde\rho_{mnvb}.
\label{deltav}
\end{equation}

The equations (\ref{lcore}) and (\ref{lval}) are very similar 
to the well-known linearized
coupled-cluster single-double (SD) equations (see, e.g.~\cite{SD}). 
However, certain terms are removed from the SD equation to arrive to
Eqs.~(\ref{lcore}) and (\ref{lval}). This is because we are going
to combine the equations with the correlation potential method and 
removal of the terms is needed to ensure no double counting.
Only terms corresponding to the ladder diagrams need to be included.

Since Brueckner energy $\epsilon_v$, in the equation (\ref{Brueck}) 
and the correction $\delta \epsilon_v$, in the equation (\ref{deltav}) 
both include the second-order correlation correction, it is convenient 
to define the correction associated with the ladder diagrams as a difference
\begin{equation}
  \delta \epsilon_v^{(l)} = \delta \epsilon_v - \langle v | \hat \Sigma^{(2)}| v \rangle.
\label{deltal}
\end{equation}
Here $\hat \Sigma^{(2)}$ is the second-order correlation potential given by 
four diagrams on Fig.~\ref{sigma1}. The correction (\ref{deltal}) is additional
to the corrections considered in previous section. 

\begin{figure*}
\centering
\epsfig{figure=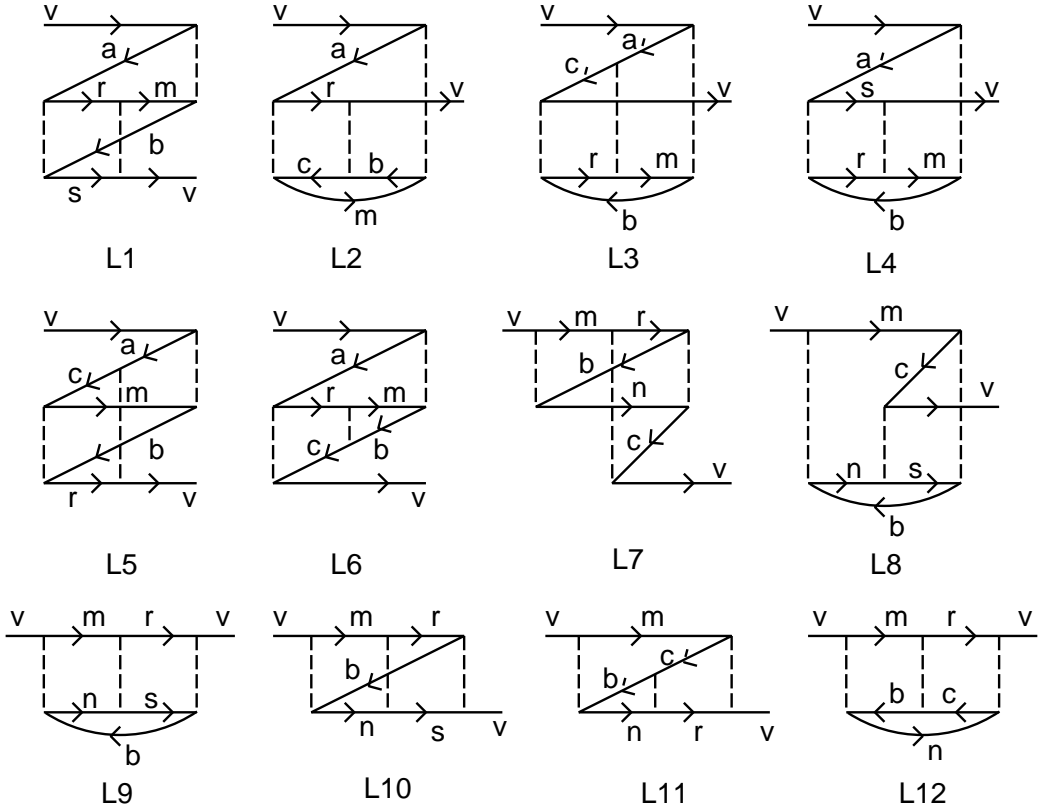,scale=0.8}
\caption{Third order ladder diagrams.}
\label{lad3}
\end{figure*}

If equations (\ref{lcore}) and (\ref{lval}) are iterated simultaneously, i.e. 
one iteration is done for Eq.~(\ref{lcore}) and one iteration is done 
for Eq.~(\ref{lval}), and then process is repeated again, etc., 
then every iteration corresponds to the next
order of the MBPT. For example, single iteration of both sets of equations
produces the third-order ladder diagrams. All third-order ladder diagrams are shown 
on Fig.~\ref{lad3}. Comparison with the 
third order correlation potential diagrams presented on Fig.~\ref{cor3}
shows that there is no overlap between them. This means that ladder
diagrams represent a new class of the higher-order diagrams which was
not included into the correlation potential method.

The results of the calculations
of the ladder correction $\delta \epsilon_v^{(l)}$ for the lowest $s,p$ and $d$
states of Cs and Tl are presented in Table~\ref{tb:lad}. Comparison with the 
data from Table~\ref{tb:CP} shows that inclusion of the ladder diagrams leads
to significant improvements of the accuracy of the results in practically
all cases. However, these results are not final since we are going to
include ladder diagrams into the correlation potential $\hat \Sigma$. 
This would slightly change the results. Apart from that, small difference
between theory and experiment after inclusion of the ladder diagrams
means that other small corrections such as Breit interaction and
quantum electrodynamic corrections (QED) need to be considered as well.
This will be done in next section.

\begin{table}
\caption{Ladder diagrams corrections to the energies of the lowest
$s,p$ and $d$ states of Cs and Tl (cm$^{-1}$).}
\label{tb:lad}
\begin{ruledtabular}
  \begin{tabular}{lrlr}
\multicolumn{2}{c}{Cs} &\multicolumn{2}{c}{Tl} \\
\hline
State & $\delta \epsilon_v^{(l)}$ & State & $\delta \epsilon_v^{(l)}$ \\
\hline
$6s_{1/2}$ & -131 & $7s_{1/2}$ &   -43 \\
$6p_{1/2}$ &  -60 & $6p_{1/2}$ & -1215 \\
$6p_{3/2}$ &  -54 & $6p_{3/2}$ &  -794 \\
$5d_{3/2}$ & -189 & $6d_{3/2}$ &   -29 \\
$5d_{5/2}$ & -193 & $6d_{5/2}$ &   -27 \\
  \end{tabular}
\end{ruledtabular}
\end{table}

\section{Calculations for cesium and thallium}

It has been demonstrated in previous section that the calculations
of the energy levels of such atoms as cesium and thallium by means of the
all-order correlation potential method can be significantly improved
if contributions of the ladder diagrams are also included.
The remaining discrepancy between theory and experiment is a fraction
of a per cent for all low $s$, $p$ and $d$ states of both atoms. 
This means that other small corrections to the energy need 
to be considered.

In this section we consider inclusion of the ladder diagrams into the
correlation potential $\hat \Sigma$ as well as the Breit and QED
corrections. 

\subsection{Inclusion of the ladder diagrams into the correlation potential}

For accurate calculations of the matrix elements its is important to 
have as accurate correlation potential $\hat \Sigma$ as possible.
This would allow to have accurate Brueckner orbitals by solving
the equations (\ref{Brueck}) for valence electrons. It would also lead 
to more accurate values of such corrections to the matrix elements as 
structure radiation and renormalization which can also be expressed in 
terms of $\hat \Sigma$~\cite{CPM}.
Therefore, it is important to include ladder diagrams into the
correlation potential $\hat \Sigma$.
This is done by modifying the expressions for the second-order
correlation potential $\hat \Sigma^{(2)}$ (Fig.~\ref{sigma1}).
Each term for the diagrams 1 and 2 on Fig.~\ref{sigma1}
is multiplied by the factor 
\[  \frac{\rho_{vanm}}{g_{vanm}(\epsilon_v+\epsilon_a-\epsilon_n-\epsilon_m)}-1, \]
and each term for the diagrams 3 and 4 is multiplied by a similar factor 
\[  \frac{\rho_{vabm}}{g_{vabm}(\epsilon_a+\epsilon_b-\epsilon_v-\epsilon_m)}-1, \]
The meaning of $\rho,g,\epsilon$ and all indexes is the same as in equation
(\ref{lcore}) and (\ref{lval}). Subtraction of one is needed to exclude double 
counting of the second-order correlation potential $\hat \Sigma^{(2)}$ (as
in (\ref{deltal})).

The results of the calculations of the energy levels of Cs and Tl
with the all-order $\hat \Sigma$ which also includes ladder diagrams
presented in Table~\ref{tb:final} together with the Breit and QED
corrections.

\subsection{Breit and QED corrections}

The Breit interaction accounts for magnetic and retardation corrections to the 
non-relativistic Coulomb interaction between atomic electrons.
We use the following form for the Breit operator,
\begin{equation}
\hat H^{B}=-\frac{\mbox{\boldmath$\alpha$}_{1}\cdot \mbox{\boldmath$\alpha$}_{2}+
(\mbox{\boldmath$\alpha$}_{1}\cdot {\bf n})
(\mbox{\boldmath$\alpha$}_{2}\cdot {\bf n})}{2r} \ ,
\label{Breit}
\end{equation}  
where ${\bf r}={\bf n}r$, $r$ is the distance between electrons, and 
$\mbox{\boldmath$\alpha$}$ is the Dirac matrix.

In a similar way to the Coulomb interaction, we determine the self-consistent 
Hartree-Fock contribution arising from Breit. This is found by solving 
the Hartree-Fock equations for single-electron orbitals in the potential
\begin{equation}
\hat V=V^{C}+V^{B} \ ,
\end{equation}  
where $V^{C}$ is the Coulomb potential, $V^B$ is the Breit potential.
Coulomb interaction in the second-order $\hat \Sigma$ (Fig.~\ref{sigma1})
is also modified to include Breit operator (\ref{Breit}).
The Breit correction to the energy of external electron is found
by comparing the second-order Brueckner energies (Eq.~(\ref{Brueck})) 
calculated with and without Breit interaction.

Quantum electrodynamics radiative corrections to the energies (Lamb shifts) 
are accounted for by use of the 
radiative potential introduced in Ref. \cite{radpot}. This potential has the form
\begin{equation}
V_{\rm rad}(r)=V_U(r)+V_g(r)+ V_e(r) \ ,
\end{equation}
where $V_U$ is the Uehling potential, $V_g$ is the potential arising from the 
magnetic formfactor, and $V_e$ is the potential arising from the electric formfactor.
As for the case of Breit interaction, the QED corrections to the energies
of external electron are found by solving equations (\ref{Brueck}) with
and without radiative potential.

The results for the Breit and QED corrections are presented in Table~\ref{tb:final}.

\subsection{Discussion}

Final results of the calculations of the energy levels of cesium
and thallium are presented in Table~\ref{tb:final}. They include 
the all-order correlations considered in section \ref{CPM}, ladder diagrams
included into the correlation potential $\hat \Sigma$, Breit and
QED corrections. The difference between theory and experiment ($\Delta$)
is a fraction of a per cent in all cases. This represents significant 
improvement comparing to the correlation potential method considered
in section \ref{CPM}.

It is important that the accuracy is now about the same for
all states of both atoms. This proves the claim that poor
accuracy for $d$ states of cesium and $p$ states of thallium
in the all-order correlation potential method (section \ref{CPM})
is due to the poor treatment of the residual Coulomb interaction
between external electron and the core. 
Inclusion of this effect via ladder diagrams leads to significant 
improvement for these states while it has little effect on the
states where accuracy is already high.

In the case of thallium the main source of uncertainty is the
choice of screening factors for calculation of the exchange
diagrams (Table~\ref{tb:fk}). Relative contribution of the
exchange correlation diagrams for thallium is larger than 
for cesium. Therefore, approximate inclusion of higher-order
correlations into exchange diagrams via the use of screening
factors works very well for cesium but not so well for thallium.
One possible cause of action for further improvement is the use
of the Feynman diagram technique for the exchange diagrams as 
well as for direct diagrams as it was done in our calculations 
of the PNC in cesium~\cite{Dzuba02}.

\begin{table}
\caption{Final results for the energies of the lowest $s$, $p$ and $d$ 
states of Cs and Tl (cm$^{-1}$); comparison with experiment.}
\label{tb:final}
\begin{ruledtabular}
  \begin{tabular}{ccc rrrrr}
Atom & State    & \multicolumn{1}{c}{$\hat \Sigma$\footnotemark[1]} & 
\multicolumn{1}{c}{Breit} & 
\multicolumn{1}{c}{QED} & 
\multicolumn{1}{c}{Sum} & 
\multicolumn{1}{c}{$\Delta$\footnotemark[2]} &
\multicolumn{1}{c}{Exp.~\cite{Moore}} \\ 
\hline
Cs & $6s_{1/2}$ & 31402	&   -4  & -22  & 31376 &   31 & 31407 \\
   & $6p_{1/2}$ & 20191	&  -10  &   1  & 20182 &   47 & 20229 \\
   & $6p_{3/2}$ & 19632	&   -4  &   0  & 19628 &   47 & 19675 \\
   & $5d_{3/2}$ & 16901	&   20  &   5  & 16926 &  -18 & 16908 \\
   & $5d_{5/2}$ & 16814	&   21  &   4  & 16839 &  -29 & 16810 \\
	       		 	 	      	    		 	 	 	  
Tl & $7s_{1/2}$ & 22943 &  -26  & -24  & 22893 & -107 & 22786 \\
   & $6p_{1/2}$ & 49466 & -251  &  38  & 49253 &   11 & 49264 \\
   & $6p_{3/2}$ & 41613 & -126  &  29  & 41516 &  -45 & 41471 \\
   & $6d_{3/2}$ & 13239 &   -7  &   3  & 13235 &  -89 & 13146 \\
   & $6d_{5/2}$ & 13110 &   -5  &   3  & 13108 &  -44 & 13064 \\
    \end{tabular}
\footnotetext[1]{Brueckner orbitals with the all-order $\hat \Sigma$ 
including ladder diagrams.}
\footnotetext[2]{$\Delta = E_{\rm exp} - E_{\rm calc}$}
\end{ruledtabular}
\end{table}

The efficiency of the present method of calculations is the same as for
the standard SD approximations. Although the equations (\ref{lcore})
and (\ref{lval}) have fewer terms than the SD equation the calculations
are strongly dominated by a singe term which has double 
summation over states above the core. This term exists in both cases, 
the equations (\ref{lcore}) and (\ref{lval}) for ladder diagrams and
in the standard SD equations. Computer resources needed for the 
calculation of the all-order correlation potential $\hat \Sigma$ are 
practically negligible compared to the iterations of the equations
(\ref{lcore}) and (\ref{lval}). It would be correct to say that the
use of the all-order correlation potential method in combination with
the equations similar to the SD equations is a way to include important
triple and higher excitations without affecting the efficiency of the
calculations. Another way of describing the method is to state that
it adds one more class of the higher-order correlations to the all-order
correlation potential method. This class represents residual
Coulomb interaction of the external electron with atomic core.

\section{conclusion}

The all-order correlation potential method is extended to include one
more class of higher-order diagrams to all orders. This class describes
residual Coulomb interaction of an external electron with atomic core and
represented by ladder diagrams. This is in addition to such higher-order
effects as screening of Coulomb interaction between atomic electrons by
core electrons, interaction between an electron excited from the core with 
a hole created by this excitation and the iterations of the correlation
operator. Calculations of the energy levels of cesium and thallium show 
significant improvement in the accuracy. This opens a way of more
accurate calculations for many important applications.

\section*{Acknowledgments}

The work was supported in part by the Australian Research Council.

\end{document}